# Comment on "High energy neutron scattering from hydrogen using a direct geometry spectrometer", C Stock, R A Cowley, J W Taylor and S. M. Bennington.

**J Mayers[1], N I Gidopoulos[1], M A Adams[1], G Reiter[2], C Andreani[3] and R Senesi[3]**

[1] ISIS facility, Rutherford Appleton Laboratory, Chilton, Didcot, UK OX110QX

[2] Physics Department and Texas Centre for Superconductivity, University of Houston, Houston, Texas 77204-5005

[3] Università degli Studi di Roma Tor Vergata, Dipartimento di Fisica and Centro NAST (Nanoscienze-Nanotecnologie-Strumentazione), via della Ricerca Scientifica 1, 00133 Roma, Italy

**Abstract**

The paper in the title [1] reports measurements of neutron scattering from hydrogen in the 1-100 eV range of energy transfers, using the direct geometry MARI spectrometer at ISIS. Stock et al claim that their measurements have better or comparable energy resolution to those on the inverse geometry VESUVIO spectrometer at ISIS. Most importantly the main conclusions of ref [1] are not valid unless this claim is true: in particular the conclusion that anomalous neutron cross sections measured on VESUVIO [2] are "*the result of experimental issues using indirect geometry spectrometers*". We present here overwhelming evidence that the energy resolution of the measurements in ref [1] is much coarser than on VESUVIO. It follows that the conclusions of Stock et al are unfounded. In reality the measurements of reference [1] serve mainly to demonstrate that at eV neutron energies, direct geometry chopper spectrometers have greatly inferior energy resolution to inverse geometry spectrometers based on resonance foil methods.

## 1. Introduction

There have been many previous neutron measurements on hydrogen at eV energy transfers using inverse geometry methods [3], but the measurements reported in ref [1], are the first using direct geometry. Such measurements are indeed a welcome development. Unfortunately ref [1] is very misleading. Stock et al have withdrawn some of the incorrect claims presented in earlier versions [4] of this paper. However it is still claimed that hydrogen measurements on the VESUVIO inverse geometry spectrometer at ISIS, have worse or at best comparable energy resolution to the measurements of ref [1] on the MARI spectrometer at ISIS.

In reality, the energy resolution of VESUVIO is always considerably better than that of the measurements in reference [1]. At all but the highest scattering angles and lowest incident energies, the resolution of VESUVIO is between one and three orders of magnitude better. This statement holds true for any sample including those containing hydrogen. This implies that the measurements in ref [1] provide no basis for the claim of Stock et al that that anomalous neutron cross sections measured on VESUVIO [2] are "*the result of experimental issues using indirect geometry spectrometers*".

## 2. Comparison of instrumental resolution at MARI and VESUVIO

### 2a Heavy atoms

Stock et al accept that the "*intrinsic experimental resolution*" of VESUVIO is better than that of MARI for scattering from heavy atoms such as lead or vanadium: they state *"Such measurements find the energy widths to be narrower on Vesuvio in comparison to MARI."* It is worthwhile to investigate how superior the VESUVIO resolution for lead is compared to MARI. Fig. 1 shows lead data measured on VESUVIO. The spectra were obtained by converting time of flight spectra at constant angle to energy transfer, using the known final energy and standard methods [5]. This should be compared with the lead data displayed in Fig. 7 of ref [1]. The full width at half maximum (FWHM) of lead peaks on VESUVIO is ~0.25 eV compared with between 4 and 55 eV on MARI.

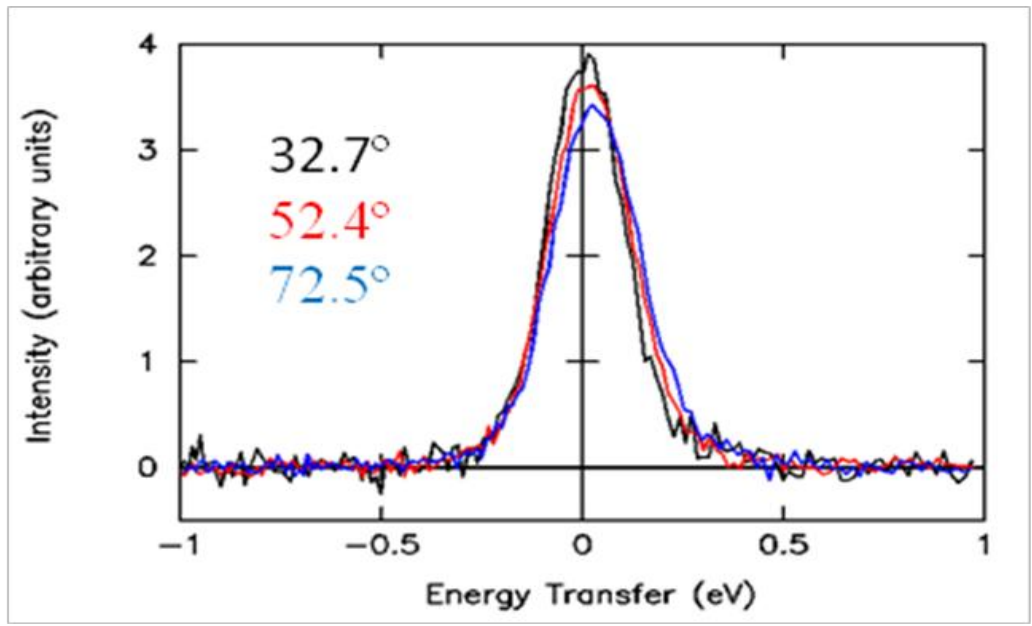

Figure. 1. Scattering from 2 mm of lead on VESUVIO at different scattering angles. The slight shift of the peak with angle is due to the increasing recoil of the lead atoms as the angle increases.

Fig. 2 shows Gaussians with the same FWHM as the lead peaks measured on VESUVIO and MARI. The lead peaks widths are almost equal to the energy resolution widths close to zero energy transfer. After taking into account the momentum distribution of the lead atoms, the energy resolution function on VESUVIO has a FWHM of ~0.2 eV [6] compared with between 4 and 55 eV on MARI. The VESUVIO energy resolution close to the elastic line is thus between 20 and 300 times better than that on MARI.

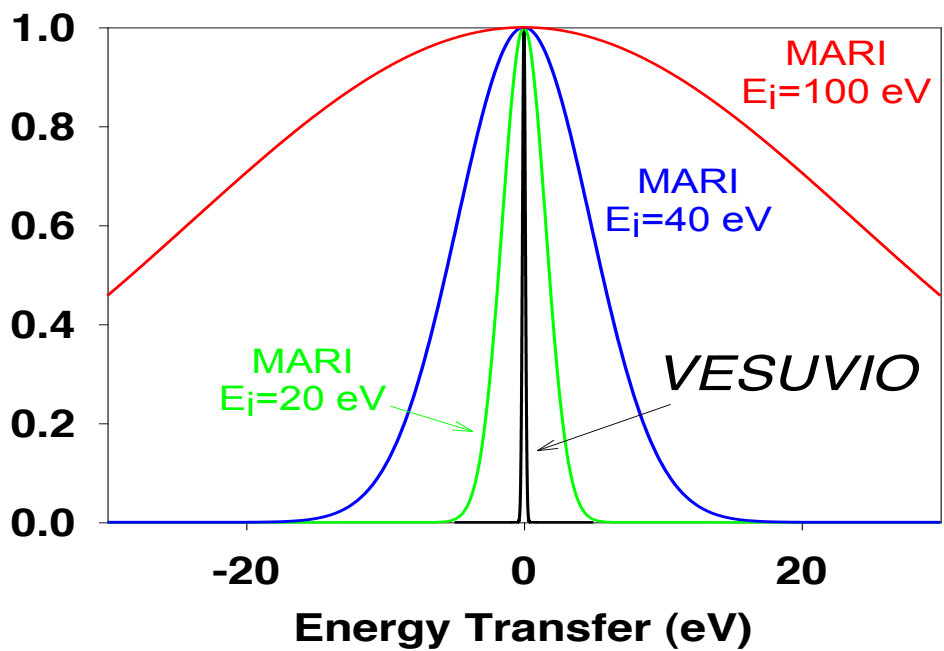

Figure 2. Gaussians with the same FWHM in energy transfer as the lead peaks measured on VESUVIO and MARI. The widths (FWHM) of the MARI data were taken from Fig. 7 of ref [1]. To a good approximation these curves show the energy resolution functions of the two instruments close to zero energy transfer.

**2b Hydrogen**

Despite the much better resolution of VESUVIO for measurements of heavy atoms, Stock et al argue that, specifically for scattering from hydrogen, the energy resolution of VESUVIO is either worse or at best comparable to that of MARI. Contrary to their claims, the much coarser energy resolution of MARI is also clearly demonstrated by MARI and VESUVIO measurements of hydrogen peak widths.

The most basic method of analysing hydrogen data on VESUVIO is to assume that the dynamic structure factor $S(Q, E)$ has the form

$$S(Q, E) = \frac{M}{\hbar^2 Q \sigma} \frac{1}{\sqrt{2\pi}} \exp\left[\frac{-y^2}{2\sigma^2}\right] \qquad (1)$$

where

$$y = \frac{M}{\hbar^2 Q}\left(E - \frac{\hbar^2 Q^2}{2M}\right) \qquad (2)$$

$Q$ is the wave-vector transfer $E$ is the energy transfer and $M$ is the proton mass. Eqs (1)-(2) are exactly correct if the Impulse Approximation (IA) is valid and the binding is by isotropic harmonic forces [7]. We note that eqs (1)-(2) are also implicitly assumed to be true by Stock et al in their eq (6).

Eq (2) is a more general consequence of the IA and implies that the variables $Q$ and $E$ are not independent. This is known as "y scaling", and is rigorously accurate in any sample if $Q$ is sufficiently large [8]. A consequence of y scaling, which is fundamental to all data analysis on VESUVIO, is that for samples with no preferred direction (eg liquids, powders) any scan in ($Q, E$) space which crosses the recoil line, $E = \frac{1}{2} Q^2 / M$, gives identical information. Thus the hydrogen peak in spectra at all angles can be fitted with the same two parameters; $\sigma$ and the peak amplitude. $\hbar\sigma$ is physically the root mean square momentum of protons in the sample. The peak amplitude determines the scattering intensity and hence the neutron cross-section. More details of exactly how this procedure is implemented on VESUVIO are given in ref [9].

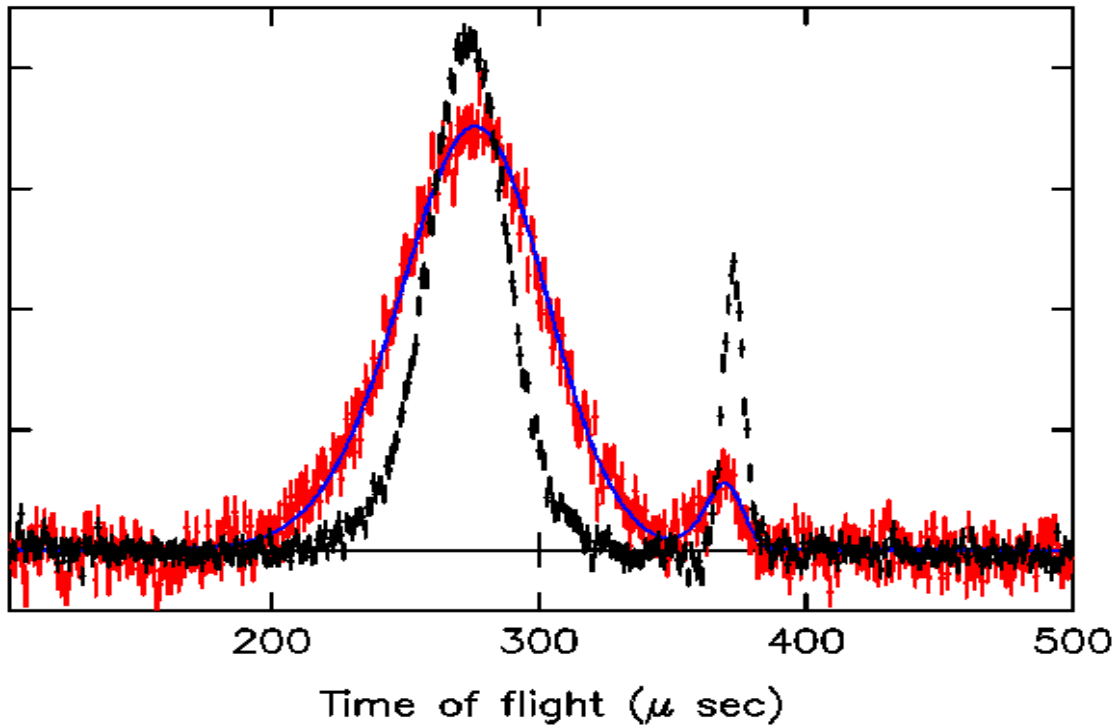

Figure 3. Uncorrected time of flight data from $CH_2$ and liquid $H_2$ on VESUVIO. The black points were taken from a sample of liquid hydrogen contained in an aluminium can. The red points are data from $CH_2$. The blue line is the fit of eqs

(1) and (2) to the hydrogen and carbon peaks in the CH$_2$ data. The data was collected in a detector at a scattering angle of 45.9º. The intensity is normalised to the incident beam monitor and is in arbitrary units.

The blue line in Figure 3 shows a typical fit of eqs (1) and (2) to the carbon and hydrogen peaks in CH$_2$ data collected on VESUVIO. The data, shown as the red points with error bars, is uncorrected for multiple scattering and background. Figure 4 shows values of $\sigma$, as a function of scattering angle, obtained from fits of eqs (1)-(2) to the hydrogen peak in CH$_2$ data after correction for the latter two (small) effects. The black points are obtained after further correction for the calibrated [6,10] energy resolution width and for deviations from the IA due to the finite $Q$ of the measurements as described in ref [9]. The red points were obtained with no correction for the energy resolution, the blue points with no correction for deviations from the IA. The statistical error bars (due to counting statistics) are shown as the black vertical lines.

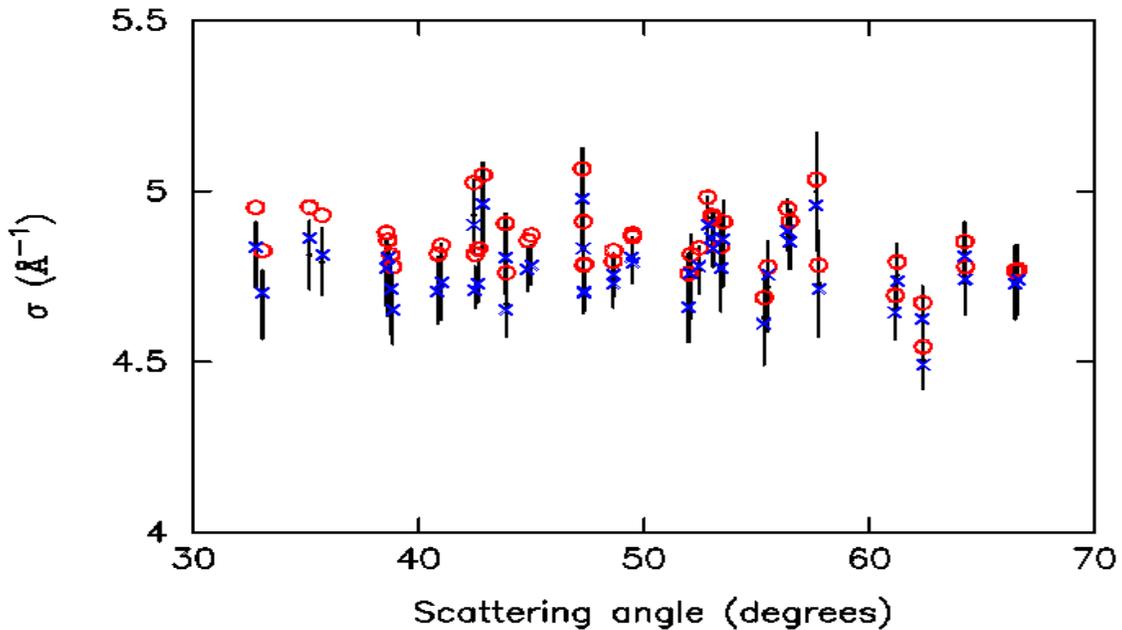

Figure 4, Values of $\sigma$ for the hydrogen peak in CH$_2$ measured on VESUVIO as a function of scattering angle, are shown as the black points. The statistical error bars due to counting statistics are shown as vertical black lines. The red circles are obtained if no correction is made for the VESUVIO energy resolution. The blue crosses were obtained with no correction for inaccuracies in the IA.

Table 1 shows the mean of the measured values of $\sigma$ displayed in Figure 4. The second column gives the mean weighted by the statistical error, with the statistical error in the mean. Column 3 gives the mean and the standard error in the mean. These were obtained using standard formulae [11].

|  | Weighted mean ($Å^{-1}$) | Mean and standard error ($Å^{-1}$) |
| --- | --- | --- |
| Measured value | 4.766 ±0.012 | 4.770 ±0.014 |
| No resolution correction | 4.844 ±0.012 | 4.848 ±0.014 |
| No FSE correction | 4.765 ±0.012 | 4.770±0.014 |

Table 1. Mean values of $\sigma$ obtained from the fitted values shown in Figure 4. The second column is the mean value weighted by the statistical error. The third column gives the mean and the standard error in the mean.

The results in Fig 4 and Table 1 demonstrate that the Impulse Approximation and hence the assumption of "y scaling" on VESUVIO is accurate for measurements on hydrogen in any sample. For example;

(i) Fig 4 shows that the fitted values of $\sigma$ are independent of angle between 32° ( $Q$~31$Å^{-1}$) and 66° ( $Q$~112$Å^{-1}$). Comparison of the mean values and errors in columns 2 and 3 of Table 1 shows that this is true almost to within the (very small) error due to counting statistics.

(ii) Values of $\sigma$ are not significantly affected by (well understood [8]) corrections for inaccuracies in the IA. This again shows that such effects are small and hence that y scaling is accurate.

(iii) Deviations from the IA are proportional to $\sigma/Q$ [12]. Since protons in $CH_2$ have larger values of $\sigma$ than most other systems, if y scaling is accurate on VESUVIO for protons in $CH_2$ it will be accurate for protons in any condensed matter system.

The quality of the fits obtained using eqs (1) and (2), (see for example Fig 3) together with the internal consistency of data collected at different scattering angles and demonstrated in Figure 4 and Table 1, implies that eqs (1) and (2) provide a very accurate description of VESUVIO $CH_2$ data. Hence it follows from eqs (1) and (2) that the widths $\sigma$ obtained from VESUVIO can be converted to the width $2\Gamma$ of the hydrogen peak at constant $Q$ via,

$$2\Gamma = 2\sqrt{2\ln 2}\frac{\hbar^2 \sigma Q}{M} = 0.00977\, Q\sigma \qquad (3)$$

where the second equality applies if $2\Gamma$ is in eV and $Q$ and $\sigma$ are in Å$^{-1}$. For example, at $Q$ =100 Å$^{-1}$, and $\sigma = 4.844$ Å$^{-1}$, we obtain $2\Gamma$ =4.73 eV, close to the value of ~5 eV calculated from Vesuvio CH$_2$ data in ref [1], by use of the Waller-Froman factor.

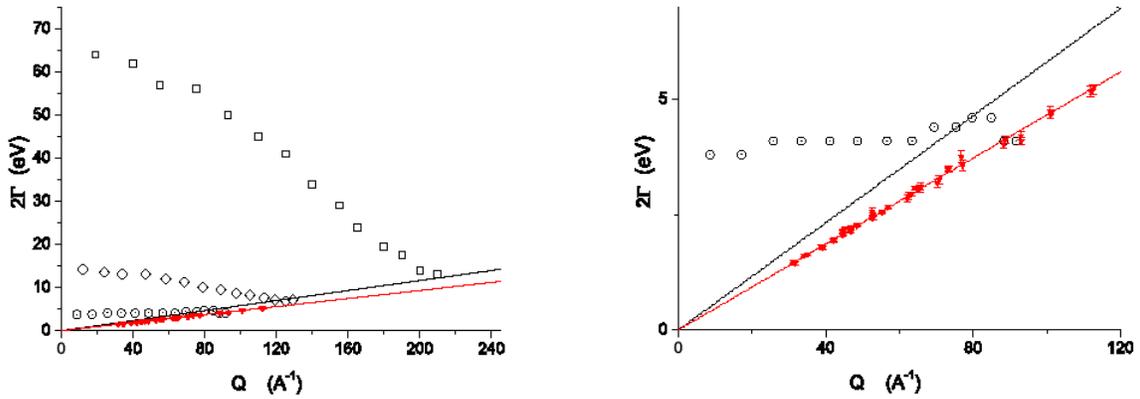

Figure 5. The squares are values of $2\Gamma$ obtained in ref [1] with incident energies of E$_0$ =100 eV. The diamonds are the values obtained with E$_0$ =40 eV and the circles with E$_0$ =20 eV. The red points are VESUVIO data. The red line is the best fit to the VESUVIO data points (the fitted line crosses the origin without enforcing this). The same line is also that obtained from eq (3) with $\sigma$ = 4.77 Å$^{-1}$. The line without any resolution correction ($\sigma$ = 4.84 Å$^{-1}$) is indistinguishable from the red line. The black solid line is the line passing from the origin which best fits the three MARI data points that correspond to minimum widths for each incident energy. If crossing the origin is not enforced, the fitted black line would cross the x axis at Q = 40 Å$^{-1}$. The right hand figure shows the same data on an expanded energy scale.

The values of $2\Gamma$ obtained on VESUVIO, from eq (3) and the values of $\sigma$ in Fig 4, are shown in Figure 5 as the red points. The $Q$ of each measurement was taken as the value where the constant angle scan crosses the recoil line. The values of $2\Gamma$ shown in Figure 13 of ref [1] are also displayed. It is obvious that in almost all $Q, E$ space $2\Gamma$ is much larger on MARI than on VESUVIO. This can only be due to the much coarser resolution of MARI. To a good approximation $2\Gamma$ is determined by adding in quadrature the instrument energy resolution width $W_R$ and the "*intrinsic*" peak width $W_H$, due to the momentum distribution of the protons.

$$2\Gamma = \sqrt{W_R^2 + W_H^2} \qquad (4)$$

For most of the MARI points shown $W_R >> W_H$ and $2\Gamma$ is determined almost entirely by the instrument energy resolution. On VESUVIO it is always the case that $W_R << W_H$ and $2\Gamma$ is determined almost entirely by the sample response.

The measurements in figures 1-5 are conclusive evidence that at eV energies, the energy resolution of VESUVIO is greatly superior to that of MARI for both hydrogen and lead. Nevertheless, Stock et al still claim that the measurements of hydrogen in ref [1] have better or at worst comparable energy resolution to those on VESUVIO. They state in section VI that: *"the energy widths for MARI (for a fixed scattering angle 2θ) at large scattering angles is ΔE/E ~ 20%. The experimental data shown in Fig. 1 (VESUVIO) gives an energy width (at a fixed scattering angle) of about 2Δt/t ~ 50%. A comparable data set for polyethylene on Vesuvio (Ref. 35) has been analyzed and the energy widths at fixed scattering angles are all substantially larger than the widths shown in Fig. 13 except at the lowest energies where the resolution in the MARI experiment is quite coarse, but could be improved by using a lower incident energy."*

They further state in section VI of ref. [1], *"We find the minimum-energy widths of both experiments are very similar. Specifically for Q=100Å$^{-1}$, experiments on polyethylene on Vesuvio have obtained an energy width of ~ 25 eV whereas we measure ~ 7 eV. If we divide by the Waller-Froman Jacobian discussed earlier, the VESUVIO width becomes 5 eV, somewhat narrower than the MARI results presented here. Therefore, depending on how the data is described, the hydrogen recoil widths measured here are comparable to studies on Vesuvio"*.

The implication of both of these statements is that because the hydrogen peak width on MARI is narrower or comparable to that on VESUVIO, the MARI energy resolution is better or comparable to that of VESUVIO. In fact Stock et al have misinterpreted in an elementary way the VESUVIO data shown in their fig 1. They correctly state, *"the broad widths of the hydrogen recoil lines on VESUVIO are the result of the detector trajectories intersecting the recoil line more tangentially in the indirect geometry setup on VESUVIO than on direct geometry machines such as MARI."* It is hard to understand why they do not draw the unavoidable conclusion: hydrogen peak widths at fixed angle on MARI and on VESUVIO are not comparable and do not convey any direct information about the relative resolutions of the two instruments.

This conclusion is also unavoidable from other considerations. As Stock et al write, *"The energy widths results partly from the instrumental resolution and partly from the intrinsic width due to the motion of the hydrogen atoms".* It can be seen from Fig 4 and Table 1 that the energy resolution contributes only ~1-2% to hydrogen peak widths on VESUVIO. It follows that the width in energy of the hydrogen peak on VESUVIO (either at constant angle or constant $Q$) conveys no quantitative information about the VESUVIO energy resolution [13].

The basic misunderstanding of Stock et al is that they have implicitly assumed that hydrogen peak widths are determined entirely by the instrument resolution width on both MARI and VESUVIO (that is that the hydrogen momentum distribution can be treated as a delta function). This is a good approximation for all the MARI data in ref [1] except that at the highest $Q$ values. It is completely wrong on VESUVIO, where the width (at both constant angle and constant $Q$) is always determined by the sample to within ~2% (see Table 1). Fig. 3 shows uncorrected time of flight data at constant angle on VESUVIO from liquid $H_2$ (black) in addition to $CH_2$ data (red). The same detectors in identical positions were used for both measurements. The different widths (and line shape) of the hydrogen peaks at ~280 μsec can only be due to the different sample responses. As is immediately obvious from the data, the momentum distribution of the protons in $CH_2$ has a FWHM about twice that in liquid $H_2$. On MARI the energy resolution is so poor that virtually no difference in the hydrogen peak width from these samples would be observed.

**4. Fitting the hydrogen peak width and MARI resolution**

In earlier versions [4] of ref [1], it was claimed that the pulse width in time of eV neutrons leaving the reflector-moderator assembly at ISIS, is an order of magnitude larger than previously thought. It was inferred that the resolution of VESUVIO is therefore much worse than assumed. Stock et al stated; "*We consider that a large contribution to this width* (our note: that is the hydrogen width at Vesuvio) *arises from the effects that we have discussed above, namely the time width of the burst for high energy neutrons.*" If this claim had been true and the pulse width at ISIS was misunderstood to the degree claimed, not only VESUVIO data but all published data from ISIS instruments would have been suspect. However this claim was untrue as Stock et al have now acknowledged. In fact it was demonstrated by measurements a decade ago [10] that the pulse width at eV energies on VESUVIO is that predicted by calculations [14].

In ref [1] it is still argued that on MARI there are two components in the pulse width of neutrons arriving at the sample; the width $\tau_0$ produced by the Fermi chopper and a second component $\tilde{\tau}$. It is still implied that VESUVIO probably has a similar double-pulse structure and consequently that the resolution is greatly inferior to that assumed in VESUVIO data analysis. In fact it is clear that their arguments are incorrect. Independently of the three different (and contradictory) discussions in the Appendix B of the three versions of the paper by Stock et al [1,4], the physical content of $\tau_0$ and $\tilde{\tau}$ is given unambiguously by eqs. (B1) and (B3) of [1]. According to these equations $\tau_0$ is determined by the width in energy of $\Delta E_0$ of the incident pulse and $\tilde{\tau}$ by the width in energy of $\Delta E_1$ of the scattered pulse. It is evident that $\Delta E_1$ cannot be statistically independent of $\Delta E_0$. Hence these two uncertainties cannot be added in quadrature as Stock et al assume. For example, if Stock et al were correct and $\tau_0, \tilde{\tau}$ were independent, then installation of a perfect chopper on MARI ($\tau_0 = 0$), would still imply very poor energy resolution.

In our view the analysis of the hydrogen peak widths in ref [1] is misguided. In reality the variation of $2\Gamma$ with $Q$ in the MARI measurements is a straightforward consequence of the fact that the MARI incident energy at eV energies is very coarsely defined by the MARI Fermi chopper. The energy resolution of a direct geometry spectrometer is [5]

$$W_R = \left(1 + \frac{L_0}{L_1} \frac{E_1^{3/2}}{E_0^{3/2}}\right) \Delta E_0 \qquad (5)$$

where $E_0$ is the incident energy $E_1$ the final energy, $L_0$ is the incident flight path, $L_1$ the final flight path and $\Delta E_0$ is the spread of neutron energies incident on the sample. $\Delta E_0$ is determined almost entirely by the characteristics of the MARI Fermi chopper since (as Stock et al state), "*the neutrons are emitted in a short pulse about 0.5 $\mu$ sec long that will be treated by our simplified model as instantaneous*".

Using the standard IA result that $E_1 / E_0 = \cos^2 2\theta$ at the hydrogen peak centre (see eq (1) of ref [1]) and inserting the MARI values $L_0$=11.79 m and $L_1$ = 4.02 m, eqs (4) and (5) can be fitted to the MARI data with the single fitting parameter $\Delta E_0$. The value of $\sigma$ =4.766 Å$^{-1}$ obtained from VESUVIO data was used to calculate $W_H$ from eq (3). The fits are shown as the black lines in Fig 5. The fitted values of $\Delta E_0$ are listed in table 2 for the three incident energies displayed in Figure 5. Also listed values of $\Delta E_0$ obtained from the FWHM of the lead peaks shown in Figure 7 of ref [1].

| Incident energy (eV) | $\Delta E_0$ (eV) from H | $\Delta E_0$ (eV) from Pb |
|---|---|---|
| 100 | 15.9±0.4 | 14 |
| 40 | 3.63±0.03 | 2.8 |
| 20 | 1.08±0.04 | 0.9 |

Table 2. The second column gives t $\Delta E_0$, the range of incident neutron energies obtained by fitting hydrogen data as described in the text. The third column gives the FWHM's of the lead peaks shown in Fig 7 of ref [1].

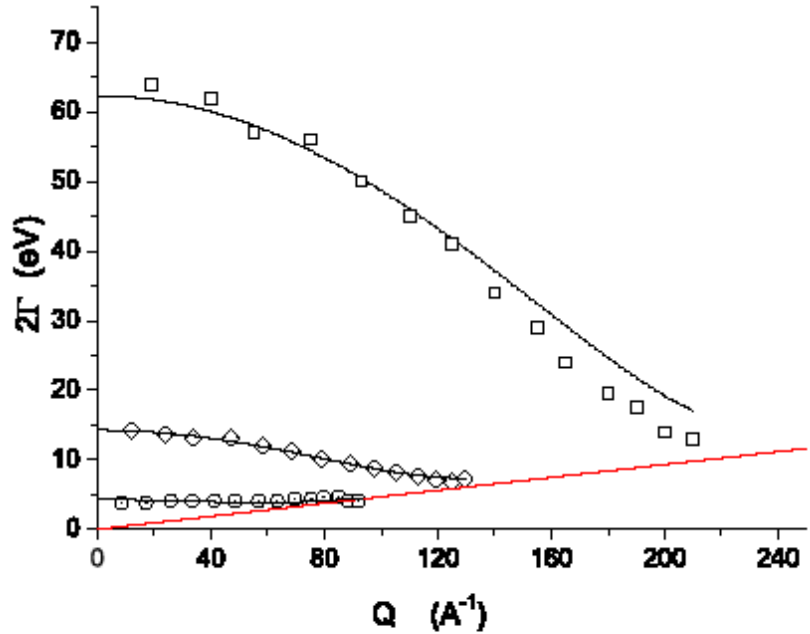

Figure 6. Data points and red solid line as in Figure 5. The black solid lines are the best fits to MARI data obtained from the single-parameter fit to eqs (4 and (5)).

It can be seen from Fig 6 that the 20 eV data is very well fitted by the single parameter $\Delta E_0$. The fit becomes slightly worse as $E_0$ increases. This is not surprising given the fitting procedures in ref [1]. Stock et al write that the values of $2\Gamma$ in Figure 13 of ref [1] were obtained by "*fitting the constant 2θ scans to the sum of two gaussians to represent the recoil lines from hydrogen and carbon* ". It appears from Figure 11 of reference [1] that the hydrogen peak shape in constant 2θ scans becomes progressively less Gaussian as $E_0$ increases. It seems very different from a Gaussian for $E_0$=100 eV. The same is true for the lead peak shape in Figure 7 of ref [1]. This change in peak shape is probably due to the increasing transparency of the MARI Fermi chopper as the incident energy is increased and implies that their fitting procedure cannot yield accurate values of $2\Gamma$.

It can also be seen from Table 2 that that the values of $\Delta E_0$ obtained from lead and hydrogen data are in quite good agreement. In fact the $Q$ dependence of $2\Gamma$ in the MARI measurements can be explained rather well by eqs (4) and (5), with $\Delta E_0$ taken from the MARI lead data. This textbook calculation gives a better overall description of the MARI hydrogen data than the model of ref [1] which includes two free fitting parameters, but cannot explain the 20 eV MARI data at all. A more accurate evaluation of the MARI $CH_2$ data could be determined by fitting the peak shape of the lead data on MARI in a similar way to that described in refs [6,10].

**5. Cross-section Anomalies**

The scientific motivation of ref [1] was the ongoing debate about the cross-section anomalies observed on VESUVIO [2,9] and by two separate groups using electron scattering [15,16]. The results in ref [1] are consistent with the measurements of Moreh et al [17,18], that with very coarse energy resolution no anomalies are observed. However, as previously pointed out [19], this is hardly surprising. Theories of anomalous cross-sections predict that there will be no anomalies with very coarse resolution.

(i) Some of the authors of this paper have other interpretations that do not agree with the arguments of Karlsson [20]. However these arguments have a clear physical basis and they have not been refuted. They rely upon a few hydrogen atoms being within the coherence volume defined by the resolution of the measurement. The coherence length of the incident beam is

$$l_0 = \lambda_0 E_0 / \Delta E_0 \qquad (7)$$

where $\lambda_0$ is the wavelength of the incident beam. The values of $l_0$ on MARI and VESUVIO at different incident energies are listed in table 4. It is clear from table 4 that while the VESUVIO coherence volume is sufficiently large to include two or more protons, that on MARI is not. Hence for the reasons given in ref [19] one would not expect any anomaly according to the theory in ref [20].

| Instrument | $\Delta E_0$ | $E_0$ (eV) | $\lambda_0$ ( Å) | $l_0$ ( Å) |
|---|---|---|---|---|
| VESUVIO | 0.3 | 100 | 0.029 | 9.5 |
| VESUVIO | 0.2 | 6 | 0.117 | 3.5 |
| MARI | 1.0 | 20 | 0.064 | 1.27 |
| MARI | 4.0 | 40 | 0.051 | 0.51 |
| MARI | 15.0 | 100 | 0.029 | 0.19 |

Table 4. Values of the coherence length $l_0$ of the incident beam in MARI and VESUVIO, obtained from eq (7).

(ii) Theories based on breakdown of the Born Oppenheimer approximation [21,22], predict that after a neutron-proton collision with sufficiently large momentum transfer Q to excite electrons non-adiabatically, neutron intensity appears at secondary peaks separated from the main recoil peak by an energy which is at least equal to the electronic gap. According to this theory, neutron intensity which would emerge at a distance from the recoil peak merges with the background and effectively "disappears". Consequently, the cross-section appears to be reduced. However, if one integrates the neutron intensity over a broad region which includes the transferred intensity, then one should expect to recover a normal rather than an anomalous cross-section.

An estimate of the amount of excitation (within a factor of ~2) is given in ref [21] The only parameters needed are the electronic energy gap and the zero point energy of the proton in its potential before collision (~140 meV). The probability that electrons are not excited (i.e. the relative cross-section) is

$$1 - 3.5134 \times 10^{-6} Q^2 \qquad (Q \text{ in Å}^{-1}). \qquad (8)$$

We first note that the predicted electronic excitation, and consequently the reduction of cross-section is small throughout the range of Q. For 20 eV incident energy the maximum amount of excitation is ~3% for Q=90. The amount of excitation is small for all Q at 40 eV incident energy (~6% for Q=135) and weak but observable for 100 eV incident energy (~14% for Q=200). Secondly, because the resolution at MARI is so broad for 40 and for 100 eV incident energies, the integration of intensity is expected to yield a normal cross-section for all Q's in the 100 eV case and a normal cross-section for all Q's except the last 2-3 data points (corresponding to high Q) for the 40 eV case. Hence it is very unlikely that the measurements in ref [1] would show any cross-section anomalies due to break down of the Born-Oppenheimer approximation.

## 6. Discussion

Stock et al state following their eq (5) that " *to a good approximation the intensity and width measured with a constant Q scan or calculated theoretically can be obtained by fitting to the spectra measured at constant scattering angle … We have chosen to fit a Gaussian profile to the spectra and this gives an excellent description of the experimental results* ". These statements seem to be contradicted by the data they show in Figures 7 and 11 of ref [1]. The peak shapes fitted are very non-Gaussian at 40 eV and 100 eV. They contain long wings which would not be fitted by a Gaussian. Unfortunately Stock et al show no examples of fits in ref [1] but it seems very unlikely that the procedure they describe would capture the true peak intensities accurately.

Even if the analysis methods in ref [1] were completely accurate, the theoretical considerations in section 5 imply that the measurements of Stock et al provide no basis for their claim that anomalous neutron cross-sections observed on VESUVIO are "*the result of experimental issues using indirect geometry spectrometers*", except possibly in the sense that only indirect geometry spectrometers have the energy resolution required to observe these anomalies.

Stock et al make a number of other incorrect claims about the supposed advantages of direct over indirect geometry spectrometers. For example their claims that; (i) "*it is impossible to subtract the short time back ground on inverse geometry machines*" (section VI). To the contrary the excellent background on VESUVIO at short times is demonstrated in the uncorrected data shown in fig 3 (particularly the better statistics $H_2$ data). (ii) "*It is not possible to vary appreciably* *the scattered neutron energy*\* on inverse geometry machines (section 2). Analyser foils of, for example, tantalum and uranium give a range of different final energies between 4 and 100 eV.

In reality the performance of inverse geometry spectrometers at eV energies is greatly superior to that of direct geometry chopper spectrometers. On VESUVIO the energy resolution is sufficiently good that not only $W_H$, but the detailed line shape of the hydrogen peak (and hence the shape of the proton momentum distribution) can be measured - see for example references [23,24,25,26,27,28,]. On MARI the energy resolution is so poor at eV energies that not even the mean kinetic energy of the protons can be determined.

Stock et al state that they wish to use chopper spectrometers to "*investigate whether neutrons can be used to study high energy magnetic and electronic excitations at energy transfers greater than ~1eV*". It is clear that they have much work to do before such studies are feasible. Small $Q$ values are required for the magnetic cross-section to be significant. This can only be achieved at high incident energies ~100eV and close to zero energy transfer. The current energy resolution of MARI under these conditions is shown in Fig 2. It is impossible to resolve magnetic peaks a few eV apart with a resolution FWHM of ~50 eV.

In contrast it has been shown [29] that inverse geometry instruments give an energy resolution of ~0.5 eV for incident energies ~100 eV: using VESUVIO, it has been possible to operate in the high-energy inelastic neutron scattering regime (HINS), resolving the OH stretching peak of ice at $\omega$ ~420 meV , with low wave vector transfer  Q< 5 Å$^{-1}$ [30].  In fact for any measurements at eV neutron energies, the energy resolution of direct geometry chopper spectrometers would have to be improved by at least two orders of magnitude to be competitive with inverse geometry spectrometers using resonance foil methods. We look forward to Stock et al demonstrating that such improvements are possible on chopper spectrometers.

This research was supported within the CNR-CCLRC Agreement No 01/9001 concerning collaboration in scientific research at the Spallation Neutron Source ISIS.

C Andreani and R Senesi acknowledge financial support of the Consiglio Nazionale delle Ricerche.

G. Reiter acknowledges the support of DOE, Office of Basic Energy Sciences under Contract No.DE-FG02- 03ER46078.

NIG acknowledges helpful discussions with R Cowley.

JM would like to thank John Tomkinson and Felix Fernandez-Alonso of the Molecular Spectroscopy Group at ISIS for their critical but constructive comments.